\documentclass[12pt]{article}

\usepackage[english]{babel}
\usepackage{graphicx}

\begin{document}

\title{\bf Can the same generation of astronomers see both the short gamma-ray bursts and their supernovae precursors?}

\author{V.I. Dokuchaev\thanks{dokuchaev@ms2.inr.ac.ru}, Yu.N. Eroshenko\thanks{eroshenko@ms2.inr.ac.ru}}

\date{}

\maketitle

{\it Institute for Nuclear Research, Russian Academy of Sciences, 
pr. 60-letiya Oktyabrya 7a, Moscow, 117319 Russia}

\begin{abstract}
We consider the possibility of observing supernova (SN) -- gamma-ray burst (GRB) pairs 
separated by several years. These energetic events correspond to the second SN explosion in a double star system followed by the formation of the very short-lived pair of compact objects and finally by the GRB generation after their merge. Possible connection of SN~1997ds and GRB~990425 is discussed.
\end{abstract}

\bigskip

It's widely believed that short GRBs, with a duration of less than $\sim2$~s, are the results of the coalescences of two neutron stars or a neutron star and a black hole in binary systems \cite{BliNovPerPol84}, \cite{Pacz86} (see \cite{Nak07} for a review). Detailed numerical simulations \cite{Rez11} have shown that jet-like structures and ultra-strong magnetic fields really appears in the merge of magnetised  neutron stars, supporting the merge model of short GRBs. These compact objects are the remnants of the core collapsed SN explosions. According to population syntheses calculations, the 2nd SN usually destroys the binary system or produces the very long-live ($\sim$ several Gyr) pair. But this statement is statistical and one can expect the existence of the short-lifetime tail in the distribution of the binary systems over their lifetimes. 

\begin{figure}[t]
\begin{center}
\includegraphics[angle=0,width=0.9\textwidth]{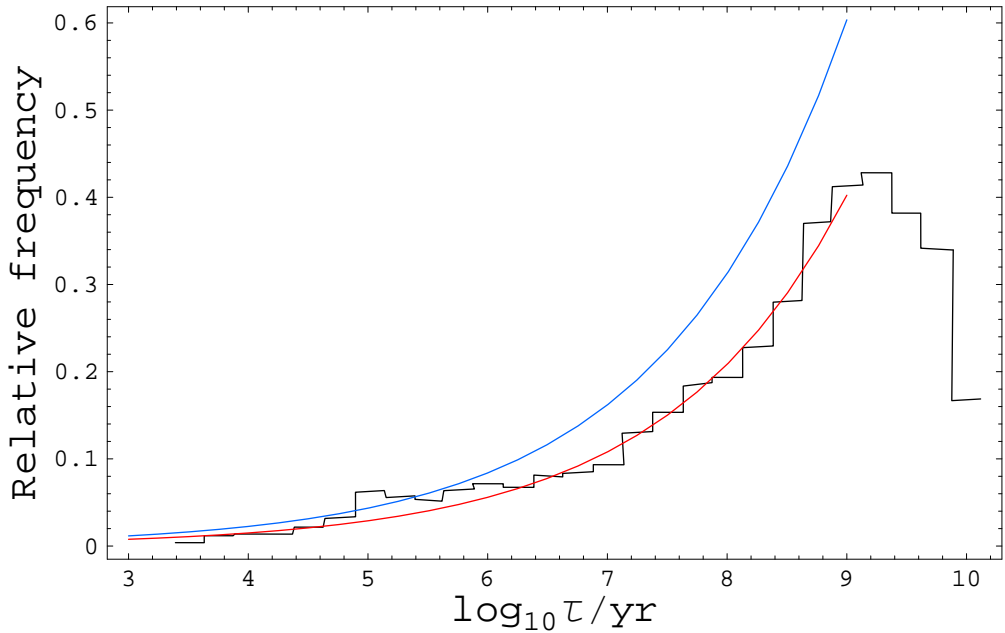}
\end{center}
\caption{The step-like curve shows the bar chart from \cite{Chu11} (low panel of Figure~4) for neutron star -- neutron star collisions lifetime with Gaussian kicks distribution. Upper smooth line corresponds to the function (\ref{prelmain}) with 0.003 coefficient, while for the bottom smooth line the coefficient 0.002 is used instead.}
\label{gr1}
\end{figure}

In the recent work \cite{We} we estimated analytically the probability that the time interval between the 2nd SN in a double system and the collision of compact objects is only several years.  
The fraction of the events, where the GRB occurs a time $<\tau$ after the
SN explosion, is 
\begin{equation}
P_{\rm rel}(<\tau)\simeq3\times10^{-3}
\left(\frac{\tau}{2\mbox{~yr}}\right)^{2/7}, \label{prelmain}
\end{equation}
where the Gaussian distribution of kick velocities was used.
This result provides the non-negligible probability of finding a GRB and its precursor SN during the several years of observations. According to (\ref{prelmain}), approximately one GRB, which is
preceded by a SN explosion, can be found among every $\simeq300$
short GRBs even in the existing astronomical catalogues. In contrast (or in addition) to an optical afterglow, such a GRB will have an optical precursor. It must be noted what the small $\tau$ limit corresponds to the non-degenerate  situation. The smallness of $\tau$ requires only the compensation of the orbital velocity by the kick velocity $v_{\rm kick}$, and the  later velocity corresponds to the typical (rather well known) region of $v_{\rm kick}$ distribution.

\begin{figure}[t]
\begin{center}
\includegraphics[angle=0,width=0.9\textwidth]{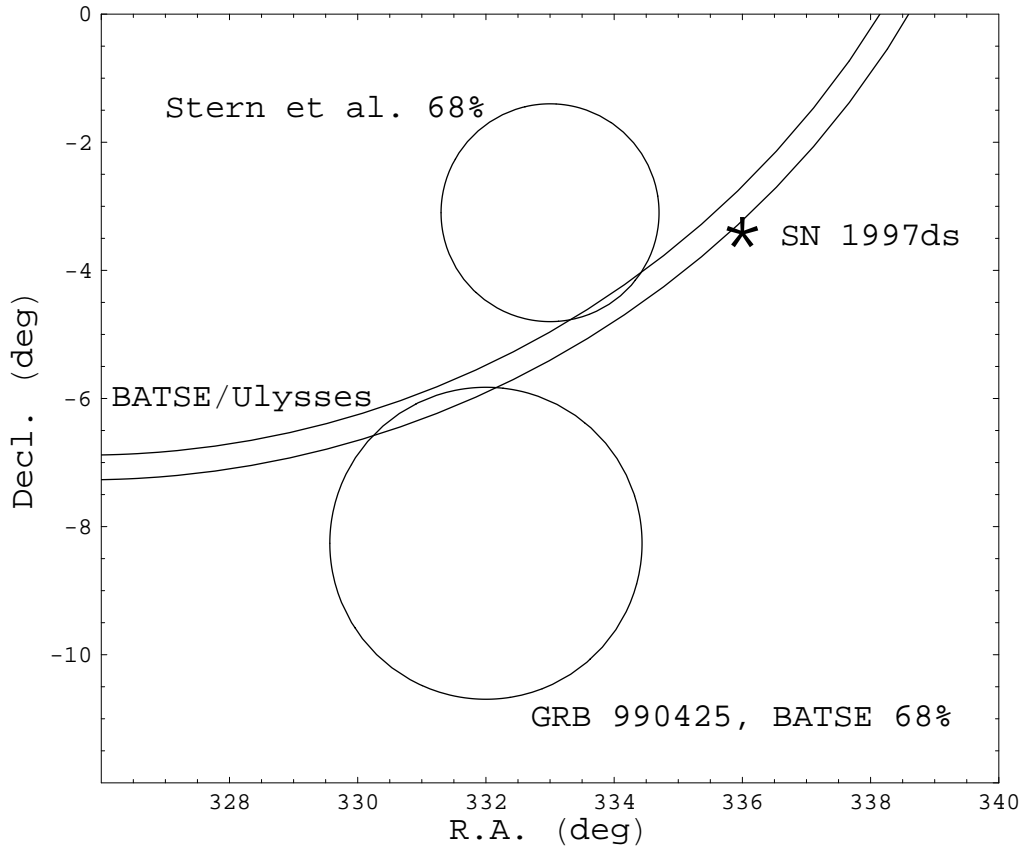}
\end{center}
\caption{The position of the SN~1997ds (marked by star) with respect to GRB~990425. Small circles show the GRB~990425 from BATSE and \cite{SteTih01} catalogues at 68\% confidence level, while the large circles show the BATSE/Ulysses triangulation annulus for this burst.}
\label{gr2}
\end{figure}

The most of calculations by the method of population syntheses are restricted by the $\tau\ge10^3$~yr 
\cite{Kal00,Chu11}. In the recent example of such a calculation \cite{Chu11} (see Fig.~4 in the paper) the distribution over the lifetimes is given. We can use even the $\tau\ge10^3$~yr
result of population syntheses \cite{Chu11} for normalization of our $P_{\rm rel}\propto \tau^{2/7}$ law from \cite{We}. For the comparison one must put $\tau=10^x$ in (\ref{prelmain}) and take derivative over $x$. The best correspondence with \cite{Chu11} calculations is obtained with the coefficient 0.002 in (\ref{prelmain}) instead of original 0.003 as it is  shown at Fig.~1. This gives the adjusted value: only $\simeq1/500$ of short GRBs must have preceding SN at the time interval of $\sim2$~yr. Our analytical curve reproduces also the result of \cite{Kal00} (see Fig.~2 in \cite{Kal00}) presented for $\tau\ge10^6$~yr. 

Then we perform the searches for the pair events in the BATSE and Swift's catalogues.
One possible coincidence was found: the SN~1997ds (R.A. $\alpha=$22h 24m 11s.51 $=336^\circ$, 
Decl. $\delta=$-03$^\circ$29'10".5) may be the two year precursor of GRB~990425 
(trigger number 11293), if one applies the GRB localization from catalogue \cite{SteTih01}.  In the \cite{SteTih01} catalogue the GRB~990425 has $\alpha=333.0^\circ$, $\delta=-3.1^\circ$ with location-error at $1\sigma$ level $r_\sigma=1.7^\circ$. Therefore the 1997ds falls into $1.8r_\sigma$ area around GRB~990425. And 1997ds lies very near the BATSE/Ulysses triangulation annulus. For the cross-correlation of the catalogues we select the core collapsed SNs and we use only 32 short and intermediate duration ($T_{90}\le3$~s) GRBs from BATSE catalogue with localisations better then $3^\circ$. A chance projection of an independent SN into the error boxes of
the above subset of BATSE GRBs is estimated on the level of $\sim1$. Therefore, at the current stage we can not state definitely the connection between SN~1997ds and GRB~990425.  The localisations of GRBs depend on the method used. For example, in the original BATSE catalogue the same GRB~990425 has $\alpha=332.67^\circ$, $\delta=-7.41^\circ$, i.e. the $\delta$ differs by about $4^\circ$ from \cite{SteTih01}, see Fig.~\ref{gr2}. This puts the additional uncertainty. Coincidences of preceding  SNs with GRBs in the Swift's catalogue, where the GRBs are very well localised, are absent.

May the SN~1997ds be really the precursor of GRB~990425? This SN is in the rather nearest galaxy MCG~-01-57-007 at redshift $z=0.009450\pm0.000020$. The SN's spectrum shows the polarisation at the level $p_V=0.85\pm0.02$\%, while each 1\% of polarisation corresponds to about 30\% of explosion asphericity \cite{SN1997ds}. The aspherical explosion may result to the kick velocity, that is necessary ingredient of the calculations \cite{We}. The GRB~990425 has the duration $T_{90}=3$~s near the boundary between short and long GRBs, so it could be regarded ``long'' as well. 

Subsequent progress in the searches for the historical connections between the short GRBs and type II supernovae may come from the accurate localizations of GRBs and their redshift measurements on the basics of afterglows observations. The angle and redshift coincidence may give the definite witness of the common origin of the SN and GRB events. Note, that we discuss here the previous SNs, i.e. the SN and GRB are separated by many years. It differs from the phenomenon of simultaneous events. For example the GRB~980425 was identified with simultaneous SN98bw \cite{Pir04}. And it would be desirable to extend the population syntheses for the very small 
$\tau\sim 1-10$~yr for better predictions of such double events. 

The possibility of observing 2nd SN (with kick of the newly formed neutron star) and following short GRB, to the best of our knowledge, was first proposed in \cite{Tro10}. In contrast to \cite{We}, the \cite{Tro10} considered  the direct impact and the interaction with disk of SN debris as the main effects of orbit decay. This require rather close periastron approach after 2nd SN in comparison with the gravitational waves energy loss \cite{We}. And respectively, \cite{Tro10} have got the time delay between the SN and GRB of the order of several days and very small probability of the double event. As an alternative scenario of short GRBs with  previous SN one can point out also the phase transition in neutron stars, the possible model is also discussed in \cite{DadDarRuj}, and the several scenarios were listed in \cite{We}.

We are thank to A.~Dar and S.~Dado for useful discussion. This work was supported by the grants of the Russian Leading
scientific schools 3517.2010.2 and Russian Foundation of the Basic Research 10-02-00635.


\begin{thebibliography}{10}

\bibitem{BliNovPerPol84} S. I. Blinnikov, I. D. Novikov, T. V. Perevodchikova,
and A. G. Polnarev, {\it Pisma v Astron. Zh.} {\bf 10}, 422
(1984) [{\it Sov. Astron. Lett.} {\bf 10}, 177 (1984)].
\begin{scriptsize}\begin{footnotesize}\end{footnotesize}\end{scriptsize}
\bibitem{Pacz86}B. Paczynski,  {\it Astrophys. J} {\bf 308}, L43 (1986).

\bibitem{Nak07}E. Nakar, {\it Phys. Rept.} {\bf 442} 166 (2007); arXiv:astro-ph/0701748v2.

\bibitem{Rez11}L. Rezzolla et al., arXiv:1101.4298v2 [astro-ph.HE].

\bibitem{We}V. I. Dokuchaev and Yu. N. Eroshenko, {\it Astronomy Letters}
{\bf 37}, 83 (2011); 	arXiv:1102.4072v1 [astro-ph.HE].

\bibitem{Kal00} V. Kalogera et al., {\it Astrophys. J} {\bf 556}, 340 (2001); arXiv:astro-ph/0012038v2.

\bibitem{Chu11}R.P. Church, A.J. Levan, M.B. Davies, N. Tanvir, arXiv:1101.1088v1 [astro-ph.HE].

\bibitem{SteTih01}B.E. Stern, Ya. Tikhomirova, D. Kompaneets, R. Svensson, J. Poutanen, {\it Astrophys. J} {\bf 563}, 80 (2001); (GRB catalogue is available on-line at http://ttt.astro.su.se/groups/head/grb\_archive.html).

\bibitem{SN1997ds} D.C. Leonard, A.V. Filippenko, {\it PASP} {\bf 113}, 920 (2001); arXiv:astro-ph/0105295v1.

\bibitem{Pir04}T.~Piran, {\it Rev. Mod. Phys.} {\bf 76}, 1143 (2004); arXiv:astro-ph/0405503v1.

\bibitem{Tro10} E.~Troja et al., {\it Mon. Not. Roy. Astron. Soc.}, {\bf 401}, 1381 (2010); arXiv:0909.3632v1 [astro-ph.HE]. 

\bibitem{DadDarRuj} S.~Dado, A.~Dar and A.~De~R\'ujula, {\it Astrophys. J} {\bf 693}, 311 (2009); arXiv:0807.1962v1 [astro-ph].

\end{thebibliography}
\end{document}